\documentclass[aps, reprint, prb, twocolumn, amsmath, amssymb, groupedaddress, 10pt, longbibliography, superscriptaddress]{revtex4-1}

\usepackage{epsfig}
\usepackage{subfigure}
\usepackage{graphicx}
\usepackage{mwe}
\usepackage{duckuments}
\usepackage{amsfonts}
\usepackage[figuresright]{rotating}
\usepackage{amssymb}
\usepackage{amsmath}
\usepackage{psfrag}
\usepackage{esint} \usepackage{hyperref}
\hypersetup{colorlinks,allcolors=blue} \usepackage{multirow}
\usepackage{siunitx}
\usepackage{xcolor}
\usepackage{lipsum}
\usepackage{soul}
\usepackage[normalem]{ulem}
\usepackage{dsfont}

\global\hyphenpenalty=100000

\setcitestyle{super,open={},close={}}

\begin{document}

\title{
Nonintegral Flux Trapping in Frustrated Josephson Networks of Triplet Superconductors 
}

\author{Grayson R. Frazier}
\affiliation{Department of Physics and Astronomy, Johns Hopkins University, Baltimore, Maryland 21218, USA}
\affiliation{Kavli Institute for Theoretical Physics, University of California, Santa Barbara, CA 93106, USA}
\author{Colton Lelievre}
\affiliation{Department of Physics and Astronomy, Johns Hopkins University, Baltimore, Maryland 21218, USA}
\author{Yi Li}
\affiliation{Department of Physics and Astronomy, Johns Hopkins University, Baltimore, Maryland 21218, USA}

\date{November 2, 2026} 

\begin{abstract}
In a Josephson junction network, anisotropic coupling between spin triplet pairing correlations can lead to frustrated $d$ vector textures that support spontaneous Josephson currents and nonintegral flux trapping.
    Such networks can appear in superconducting polycrystals, as well as single-crystal superconductors.
    In analogy to classical spin systems, in which the presence of geometric frustration and anisotropic superexchange can lead to nontrivial spin textures, Josephson networks with anisotropic Josephson couplings cannot simultaneously optimize their $\mathrm{U}(1)$ superconducting phase differences and relative $d$ vector orientations.
    The internal pairing structure of Cooper pairs twists as they tunnel across the Josephson junction, and the $d$ vector texture enters as an emergent geometric phase which can spontaneously trap fractional flux.
    For real-valued Josephson couplings between unitary $d$ vectors, the emergent geometric phase reduces to a $\mathbb{Z}_2$ structure, allowing for spontaneous $\pi$-flux trapping above a critical antisymmetric coupling, distinct from usual half-quantum vortices.
    Further, we demonstrate a continuously tunable nonintegral flux for complex Josephson couplings.
The results of this work reveal new routes to engineer frustrated  Josephson networks from the interplay of magnetic textures and spin triplet superconducting pairing order.
\end{abstract}

\maketitle

\phantomsection
Frustration can arise when system geometry and interactions are incompatible, leading to unconventional ground states, emergent collective behaviour, and symmetry breaking~\cite{Toulouse1977, Vannimenus1977, Moessner2006, Balents2010, Savary2016}. 
Josephson networks provide a natural and experimentally accessible platform to study such frustrated systems~\cite{Teitel1983, Sigrist1995, Fazio2001}.
These networks consist of multiple superconducting grains which interact via Josephson coupling, and can appear in polycrystals, or as spontaneous networks in single crystal superconductors~\cite{Pekola1999, Makhlin2001, Blom2026}.
When the preferred superconducting phase differences across different junctions are mutually incompatible, the superconducting phase cannot be chosen to satisfy all Josephson junctions simultaneously, leading to spontaneous supercurrents and flux trapping~\cite{Likharev1979, Sigrist1992}.
For example, tunneling through magnetically active barriers or in presence of strong disorder can change the sign of Josephson coupling, favoring a $\pi$-phase difference, and thus forming a $\pi$-junction~\cite{Bulaevskii1977,  Spivak1991, Kirtley1997, Ryazanov2001}.
Josephson networks consisting of rings of an odd number of such $\pi$-junctions are intrinsically frustrated and can spontaneously trap half-flux quantum, \textit{i.e.} $\pi$-flux~\cite{Geshkenbein1987, Dominguez1994, Panyukov1994, Sigrist1995, Tsuei2000}.

Unconventional pairing orders, in which the pairing order transforms according to a nontrivial representation of the system symmetry, provide an ideal setting for realizing frustrated Josephson networks.
Both system geometry and pairing symmetry can induce additional phase shifts which can lead to flux trapping.
For example, in the seminal tricrystal experiments by Tsuei and Kirtley on high-$T_c$ cuprates, the interplay between $d$-wave pairing symmetry and grain orientation was shown to spontaneously trap $\pi$-flux~\cite{Tsuei1994, Kirtley1996, Tsuei2000}.
More recently, $\pi$-flux trapping has been observed in multigranular rings of time-reversal-symmetric spin triplet superconductor $\beta$-Bi$_2$Pd~\cite{Li2019, Xu2020, Zhang2025b} and
in single-crystal rings of unconventional superconductor $4$Hb-TaS$_2$~\cite{Fischer2023a, Almoalem2024a}.
Typically, frustration in Josephson networks has been understood as arising from fixed phase shifts imposed by effective tunneling, or by orbital pairing symmetry.
However, the role of the internal pairing structure of Cooper pairs in inducing frustration in Josephson networks has remained largely unexplored.

In this work, we provide a unified picture into how the internal pairing structure enters the Josephson coupling, offering a geometric perspective into the origin of frustration and flux trapping in Josephson networks.
Building on recent results demonstrating anisotropic Josephson couplings in spin triplet superconductors in proximity to noncollinear magnetic textures~\cite{Frazier2026, Frazier2026b}, we consider networks composed of spin triplet superconducting grains, in which the internal spin structure of the Cooper pair constitutes a degree of freedom.
We demonstrate that anisotropic Josephson coupling in networks of spin triplet superconductors can give rise to an emergent geometric phase from noncollinear $d$ vector textures, which can lead to nonintegral flux trapping in Josephson networks.
We examine an exemplar case of a three-grain ring and find that, above a critical value of antisymmetric coupling, the system will become frustrated, spontaneously breaking time-reversal symmetry, forming chiral $d$ vector textures, and trapping half-flux quantum.
 \phantomsection
\paragraph*{Josephson free energy and 
flux trapping---}
\addcontentsline{toc}{section}{Josephson free energy and flux trapping}

\begin{figure}
    \centering
    \includegraphics[width=\linewidth]{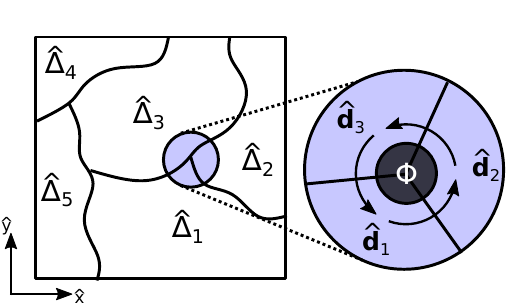}
    \caption{Schematic of Josephson junction network in a spin triplet superconductor.
    At each superconducting grain, the pairing correlation is given by $\hat{\Delta}_{n}(\mathbf{k})$, which can be expressed in terms of its $d$ vector, $\hat{\mathbf{d}}_n$.
    Between neighbouring grains, there is Josephson coupling, which is dependent on the relative orientation of $d$ vectors and $\mathrm{U}(1)$ phase difference.
    At the intersection of three grains, there is a three-grain ring, as shown on the right.
    If the Josephson network is frustrated, this can lead to spontaneous trapped flux, $\Phi$.}
    \label{fig:Josephson_network}
\end{figure}

Consider a system consisting of multiple superconducting grains, in which neighbouring grains interact via Josephson coupling, as illustrated in Fig.~\ref{fig:Josephson_network}.
In the $n\mathrm{th}$ superconducting grain, the pairing correlations are described by
\begin{math}
    \langle c_{\mathbf{k}, \alpha} c_{-\mathbf{k}, \beta}\rangle 
    \equiv
    [\hat{\Delta}_n (\mathbf{k})]_{\alpha \beta}
    =
    e^{i\varphi_n}
    [(\hat{d}_n^0 (\mathbf{k}) + \hat{\mathbf{d}}_n(\mathbf{k}) \cdot \boldsymbol{\sigma}) i \sigma^y]_{\alpha \beta}.
\end{math}
Here, 
$\varphi_n$ is the overall $\mathrm{U}(1)$ phase of the superconducting condensate,
$\mathbf{k}$ is the relative momentum of the Cooper pair about its center of mass, and $\alpha, \beta = \, \uparrow, \downarrow$ denote spin indices.
The spin singlet and spin triplet components are described by $\hat{d}_n^0(\mathbf{k})$ and the $d$ vector $\hat{\mathbf{d}}_n(\mathbf{k}) = (\hat{d}_n^x(\mathbf{k}), \hat{d}_n^y(\mathbf{k}), \hat{d}_n^z(\mathbf{k}))^\mathrm{T}$ respectively, which satisfy $\hat{d}_n^0(\mathbf{k}) = \hat{d}_n^0(-\mathbf{k})$ and $\hat{\mathbf{d}}_n(\mathbf{k}) = - \hat{\mathbf{d}}_n(-\mathbf{k})$ and in general can be complex~\cite{Balian1963, Leggett1975, Volovik2009, Mackenzie2003}.

The Josephson free energy of the network is given by
\begin{math}
    F_J
    =
    \frac{1}{2}
    \sum_{\langle n, m \rangle}
    (
    \mathcal{J}_{nm}[\hat{\Delta}_n, \hat{\Delta}_m] + \mathrm{c.c.}
    ),
\end{math}
in which the summation is taken over neighbouring superconducting grains $n$ and $m$.
Here, the first-order coupling between grains $n$ and $m$ is described by the Josephson form factor~\cite{Ambegaokar1963, Geshkenbein1986, Sigrist1991, Frazier2024},
\begin{equation}
    \mathcal{J}_{nm}[\hat{\Delta}_n, \hat{\Delta}_m]
    =
    e^{i \Delta \varphi_{nm}}
    \sum_{\alpha, \beta=0}^3
    \sum_{\mathbf{k}, \mathbf{k}'}
    \hat{d}^\alpha_n(\mathbf{k}) K_{nm}^{\alpha \beta}(\mathbf{k}, \mathbf{k}') \hat{d}_m^{\beta *} (\mathbf{k}').
    \label{josephson_form_factor.jn}
\end{equation}
Here,
the $\mathrm{U}(1)$ phase difference is given by
\begin{equation}
    \Delta \varphi_{nm} = \varphi_n - \varphi_m 
    -
    \frac{2\pi}{\Phi_0} \int \mathbf{A}(\boldsymbol{\ell})\cdot \mathrm{d}\boldsymbol{\ell},
\end{equation}
where $\Phi_0 = h/2e$ is the superconducting flux quantum of Cooper pairs, and $\mathbf{A}(\boldsymbol{\ell})$ is the electromagnetic vector potential.
We suppose the phase is uniform within each superconducting grain.
In this work, we consider the large-inductance limit for simplicity, $2\pi L I_c/ \Phi_0 \gg n$, where $L$ is the inductance, $I_c$ is the characteristic critical current, and $n$ is the number of junctions in a closed loop.
In this regime, the magnetic self-energy $\Phi^2/(2L)$ is negligible compared to the Josephson energy, and the trapped flux approaches the value determined by the geometric phase.
For finite inductance, the magnetic self-energy suppresses the spontaneous flux and can modify the preferred $d$ vector textures.

The kernel $K_{nm}(\mathbf{k}, \mathbf{k}')$ encodes the microscopic tunneling processes at the grain boundary and can be derived from, for example, linear response theory using Ambegaokar-Baratoff formalism~\cite{Ambegaokar1963, Frazier2026b}.
If the tunneling operator 
is invariant under the symmetry group, the kernel matrix $K_{nm}$ is proportional to identity, and the first order Josephson coupling is 
nonvanishing only between superconducting channels in the same irreducible representation.
In the presence of, for example, magnetic impurities or spin-orbit coupling at the junction interface, the Josephson tunneling operator transforms as a nontrivial representation of the symmetry group, allowing for Josephson coupling between different pairing channels.
These interchannel couplings are sensitive to the internal structure of the pairing order, which can induce an emergent geometric phase shift in the Josephson free energy.

The total Josephson free energy in the system can be reexpressed as
\begin{align}
    F_J
    =
    -\sum_{\langle n, m\rangle}
    |\tilde{\mathcal{J}}_{nm}|
    \cos (\Delta \varphi_{n m} + \mathcal{A}_{nm}),
    \label{free_energy.xy_analogy}
\end{align}
in which 
\begin{math}
    \tilde{\mathcal{J}}_{nm} \equiv e^{-i \Delta \varphi_{nm}}\mathcal{J}_{nm}[\hat{\Delta}_n, \hat{\Delta}_m]
\end{math}
depends only on the details of the internal orbital and spin structure of the Cooper pair.
As the Cooper pair tunnels from grain $n$ to $m$ with overall phase difference $\Delta \varphi_{nm}$, the additional phase acquired in the Josephson coupling
manifests as an emergent geometric phase,
$\mathcal{A}_{nm} = \arg (\tilde{\mathcal{J}}_{nm}) - \pi$.
For a structureless tunneling kernel, $\mathcal{A}_{nm}$ is the Pancharatnam phase corresponding to the parallel transport of the $d$ vector~\cite{Pancharatnam1956,Samuel1988},
whereas
a nontrivial kernel contributes an additional junction-dependent phase beyond the inner product of $d$ vectors, which can further frustrate the network.
The Josephson free energy between grains 
$n$ and $m$ is minimized at $\Delta \varphi_{nm} = - \mathcal{A}_{nm}$, corresponding to a $\varphi_0$-junction~\cite{Buzdin2008} for $\mathcal{A}_{nm} \neq 0 \mod 2\pi.$
For real-valued $\tilde{\mathcal{J}}_{nm}<0,$ the geometric phase is vanishing, $\mathcal{A}_{nm} = 0.$
The gauge-invariant flux in a closed loop of a Josephson junction network is given by
\begin{equation}
    \Phi
    =
    - \frac{\Phi_0}{2\pi} \sum_\mathrm{loop} 
    \Delta \varphi_{nm}
    =
    \frac{\Phi_0}{2\pi}
    \sum_\mathrm{loop}
    \mathcal{A}_{nm}.
    \label{gauge_field.flux_condition}
\end{equation}
When $\sum_\mathrm{loop} \mathcal{A}_{nm} \neq 0 \mod 2\pi$, the geometric phase cannot be removed by a global transformation.
As such, the overall superconducting $\mathrm{U}(1)$ phases become frustrated, leading to
circulating supercurrents that generate magnetic flux,
which can take on nonintegral values of $\Phi_0$.

The emergent phase $\mathcal{A}_{nm}$ originates from the internal structure of the pairing order.
In contrast to typical $\varphi_0$-junctions, where anomalous phase shifts arise from spin-orbit coupling or magnetic scattering at interfaces~\cite{Buzdin2008, Reynoso2008, Szombati2016, Amundsen2024}, the phase shift here can originate solely from $d$ vector textures.
The resulting frustration is therefore a reflection of the $d$ vector geometry of the pairing state, rather than purely from tunneling at the junction interface.
In this work, we focus on the contribution from the spin structure of the Cooper pair as it varies between superconducting grains.
The resulting geometric phase from $d$ vector textures
can produce frustrated Josephson networks and enable fractional flux trapping.

 \phantomsection
\paragraph*{Nonintegral flux trapping induced by \texorpdfstring{$d$}{d} vector textures---}
\addcontentsline{toc}{section}{Nonintegral flux trapping induced by d-vector textures}

We consider a network consisting of three superconducting grains, each with spin triplet pairing correlations, as shown in Fig.~\ref{fig:Josephson_network}.
The pairing order at grain $n$ is taken to be
\begin{equation}
    \hat{\Delta}_n (\mathbf{k}) =  
    e^{i \varphi_{n}}
    g(\mathbf{k})
    (\hat{\mathbf{d}}_n \cdot \boldsymbol{\sigma})i \sigma^y.
    \label{coarse_grained_order_parameter}
\end{equation}
in which $\varphi_n$ is the overall $\mathrm{U}(1)$ phase of grain $n$ and $\hat{\mathbf{d}}_n$ its $d$ vector orientation.
Here, the spin and orbital degrees of freedom are decoupled, with the orbital symmetry of the Cooper pair encoded in $g(\mathbf{k}),$ satisfying $g(\mathbf{k}) = - g(-\mathbf{k})$.
In this work, we are focused on the frustration arising from $d$ vector textures.
For uniform $d$ vector textures, the orbital contribution has been discussed previously~\cite{Zhang2025b}, in which a no-go theorem demonstrated that a chiral orbital pairing does not trap flux for single-band superconductors.

In addition to the $\mathrm{U}(1)$ phase difference, the Josephson form factor in Eq.~\eqref{josephson_form_factor.jn} is dependent on the relative directions of the $d$ vectors at adjacent grains~\cite{Sigrist1991, Frazier2026, Frazier2026b}.
The form factor admits the following bilinear decomposition,
\begin{equation}
    \mathcal{J}_{nm}
    =
    e^{i\Delta \varphi_{nm}}
    \Big(
    J_{nm} \hat{\mathbf{d}}_n\cdot \hat{\mathbf{d}}_m^* + 
    \mathbf{D}_{nm} \cdot  (\hat{\mathbf{d}}_n \times \hat{\mathbf{d}}_m^*)
    +
    \hat{\mathbf{d}}_n^\mathrm{T} \Gamma_{nm}\hat{\mathbf{d}}_m^*
    \Big),
    \label{dvector_couplings}
\end{equation}
corresponding to Heisenberg-like, Dzyaloshinskii--Moriya (DM)-type, and $\Gamma$-type interactions of $d$ vectors.
Here $\hat{\mathbf{d}}_n$ is the momentum-independent $d$ vector of Eq.~\eqref{coarse_grained_order_parameter}, and the momentum sums of Eq.~\eqref{josephson_form_factor.jn} are absorbed into the coefficients $J_{nm}$, $\mathbf{D}_{nm}$ and $\Gamma_{nm}$.
Such anisotropic Josephson couplings can arise in the presence of, for example, noncollinear spin textures, and have been shown to support spatially nonuniform $d$ vector textures~\cite{Frazier2026, Frazier2026b}.
For example, spin-dependent tunneling can give rise to the above Josephson couplings~\cite{Ambegaokar1963}, which follow an underlying rotation-locked structure, in analogy to the Heisenberg models studied by Shekhtman, Entin-Wohlman, and Aharony (SEA)~\cite{Shekhtman1992, Hill2021, SM}.

Remarkably, a finite geometric phase can arise even for unitary pairing states, which satisfy $\hat{\mathbf{d}}_n \times \hat{\mathbf{d}}_n^* = 0.$
Although spatial variations of unitary $d$ vector textures do not contribute to the superfluid velocity~\cite{Salomaa1987, Volovik2009, Frazier2026}, the Josephson coupling can nonetheless acquire a phase offset from the $d$ vector orientation.
For example, suppose that $\mathbf{D}_{nm} = i \mathrm{Im}\,\mathbf{D}_{nm}$ is purely imaginary and $J_{nm}$ and $\Gamma_{nm}$ are real-valued.
This is applicable to, for example, tunneling in the presence of strong $s$-$d$ coupling~\cite{Frazier2026b}.
For real-valued unitary $d$ vectors, it follows that the emergent geometric phase in Eq.~\eqref{free_energy.xy_analogy} becomes
\begin{math}
\mathcal{A}_{nm}
    =
    \arg [
    J_{nm}(\hat{\mathbf{d}}_n \cdot \hat{\mathbf{d}}_m) + \hat{\mathbf{d}}_n^\mathrm{T} \Gamma_{nm} \hat{\mathbf{d}}_m
    +
    i \mathrm{Im}\,(\mathbf{D}_{nm}) \cdot (\hat{\mathbf{d}}_n \times \hat{\mathbf{d}}_m)
    ]
    -
    \pi.
\end{math}
For a closed loop, the accumulated phase is, in general, not quantized and can trap nonintegral multiples of $\Phi_0$.
The trapped flux is dependent on the $d$ vector texture and relative magnitude of the Josephson coupling amplitudes.
For example, when $J_{nm} < 0$,
collinear $d$ vector textures do not produce additional frustration.
However, a noncollinear texture in the presence of anisotropic Josephson coupling can generate a nontrivial accumulated geometric phase.

The flux above is evaluated at a fixed $d$ vector texture, as realized when bulk anisotropic terms pin noncollinear $d$ vector textures.
Relaxation of $d$ vectors can remove this phase:
for example, for single-channel, rotation-locked couplings with a purely imaginary $\mathbf D_{nm}$ along a common axis and $J_{nm}<0$, an in-plane collinear $d$ vector texture minimizes every bond and does not trap flux, following from $|J_{nm}| \geq |\mathbf{D}_{nm}|$ for single-channel tunneling~\cite{SM}.
More broadly, whether such flux trapping survives when the $d$ vector orientations are allowed to relax remains an open question for general complex-valued couplings.

As an example, we consider an idealized three-grain ring with identical couplings $J_0 < 0$ and $\mathbf{D}_{nm} = i \tilde{D}_0 \hat{\mathbf{z}}$, with $\tilde{D}_0 \in \mathbb{R}$ at each junction.
In this toy model, 
we assume $J_0$ and $\tilde{D}_0$ can vary independently and do not follow the 
rotation-locked 
form, such that the regime $|\tilde{D}_0| > |J_0|$ can be achieved.
We impose an easy-plane anisotropy that confines real unitary $d$ vectors to the $xy$ plane, such that when $\Gamma_{nm} \propto \hat{\mathbf{z}} \hat{\mathbf{z}}^\mathrm{T}$, the $\Gamma$-type coupling vanishes.
Hence, the Josephson coupling is given by $\tilde{\mathcal{J}}_{nm} = J_0 \cos \beta_{nm} + i \tilde{D}_0 \sin \beta_{nm}$, with $\beta_{nm} \equiv \beta_m - \beta_n$ the angle between in-plane unitary $d$ vectors,
$\hat{\mathbf{d}}_n = (\cos \beta_n, \sin \beta_n, 0)^\mathrm{T}$.
Minimizing the Josephson energy over the in-plane orientations favors a collinear texture for $|\tilde{D}_0| < |J_0|$, and a chiral texture with $\beta_{nm} = \pm 2\pi/3$ for $|\tilde{D}_0| > |J_0|$.
In the latter regime, the trapped flux is given by~\cite{SM}
\begin{equation}
    \frac{\Phi}{\Phi_0}
    =
    \pm
    \left[
    \frac{3}{2\pi} \arctan
    \left(
    \frac{\sqrt{3} |\tilde{D}_0|}{|J_0|}
    \right)
    - \frac{1}{2}
    \right],
    \label{flux.three_grain.imaginary_D}
\end{equation}
which increases continuously from zero at $|\tilde{D}_0| = |J_0|$ and saturates at $\pm \Phi_0/4$ as $|\tilde{D}_0|/|J_0| \rightarrow \infty$, with the two chiral textures being energetically degenerate.
Thus, within this simple toy model, a nonintegral flux persists even when the $d$ vectors are free to relax in presence of a purely imaginary $\mathbf{D}_{nm}$.
Nonintegral flux is also possible, for example, for nonunitary $d$ vector textures, which inherently break time reversal.
If, instead,
$J_{nm}$, $\mathbf{D}_{nm}$ and $\Gamma_{nm}$ are all real-valued, 
then $\mathcal{A}_{nm}$ is restricted to $0$ or $\pi$ for unitary $d$ vectors by time-reversal symmetry.
Nonetheless,
the system can still produce $\pi$-flux from frustrated $d$ vector textures and spontaneously break time-reversal symmetry, as we demonstrate in the following section.

This distinguishes the present mechanism from those based on orbital pairing symmetry.
For time-reversal-symmetric orbital pairing order, the accumulated phase is likewise restricted to $0$ or $\pi$, such that only integer or half-integer flux is trapped.
For chiral single-band order, a finite phase is accumulated at each junction, but it depends on the grain orientations only through their differences and therefore cancels around any closed loop~\cite{Zhang2025b}.
The phase from a $d$ vector texture instead depends nonlinearly on the relative $d$ vector orientations and need not cancel; 
hence, the nonintegral trapped flux here instead arises from the internal spin structure of the pairing order.

\phantomsection
\paragraph*{\texorpdfstring{$\pi$}{pi}-flux trapping from anisotropic $d$ vector coupling---}
\addcontentsline{toc}{section}{pi-flux trapping from anisotropic d-vector coupling}

\begin{figure}
    \centering
    \includegraphics[width=\linewidth]{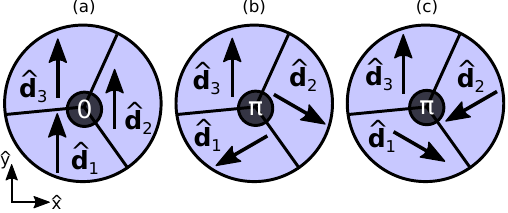}
    \caption[Minimal-energy $d$ vector configurations in the three-grain ring]{Example of minimal-energy $d$ vector configurations in the three-grain Josephson ring.
    \textbf{(a)}~For DM-like coupling $|D_0| < |J_0|/\sqrt{3}$, the $d$ vectors form a collinear texture and do not trap flux for $J_0 < 0$.
    \textbf{(b--c)}~For DM-like coupling $|D_0| > |J_0|/\sqrt{3}$, the minimal energy $d$ vector configuration is a chiral $120^\circ$-ordering, up to a sign of the $d$ vectors.
    For $D_0/J_0 < 0$, this corresponds to the chiral configuration in (b), and for $D_0/J_0 > 0$, this corresponds to the antichiral configuration in (c).
    Both chiral configurations are frustrated and spontaneously trap $\pi$-flux for $J_0<0$.
}
    \label{fig:d-vector_configurations}
\end{figure}

To demonstrate spontaneous $\pi$-flux trapping 
arising
from frustrated unitary $d$ vector textures, we analyze a minimal three-grain ring with anisotropic Josephson coupling.
We consider the case in which the bulk $d$ vectors have easy-plane anisotropy, with the easy plane normal to $\mathbf{D}_{nm}$.
As such, the $\Gamma$-type contribution from Eq.~\eqref{dvector_couplings} vanishes, leaving only Heisenberg-like and DM-like couplings,
\begin{equation}
    \mathcal{J}_{nm} = e^{i \Delta \varphi_{nm}} (J_0 \hat{\mathbf{d}}_n \cdot \hat{\mathbf{d}}_m + D_0 \hat{\mathbf{z}} \cdot (\hat{\mathbf{d}}_n \times \hat{\mathbf{d}}_m)).
    \label{J_nm.three_grain.J0_and_D0}
\end{equation}
We restrict to the case in which $J_0$ and $D_0$ are real.
Without the easy-plane confinement, 
the $d$ vector retains full $\mathrm{SO}(3)$ orientational freedom, and $\hat{\mathbf{d}}_n$ can then relieve the frustration by reorienting out of plane.
For unitary pairing, $\hat{\mathbf{d}}_n$ in Eq.~\eqref{dvector_couplings} is real up to an overall phase, which we absorb into $\varphi_n$.
The residual $\mathbb{Z}_2$ symmetry, $(\varphi_n, \hat{\mathbf{d}}_n) \equiv (\varphi_n + \pi, -\hat{\mathbf{d}}_n)$ persists.

For unitary pairing and real-valued coupling amplitudes, the emergent geometric phase $\mathcal{A}_{nm}$ reduces to a $\mathbb{Z}_2$ structure entering into the sign of the effective Josephson coupling.
The total Josephson free energy is given by
\begin{equation}
    F_J = 
    \sum_{\langle n, m\rangle}
    J^\mathrm{eff}_{nm} \cos(\Delta \varphi_{nm}),
    \label{free_energy.three_grain.real_D0}
\end{equation}
in which the effective coupling amplitude is
\begin{equation}
    J^\mathrm{eff}_{nm} = \sqrt{J_0^2 + D_0^2} \cos (\beta_{nm} + \delta),
    \label{effective_josephson_coupling}
\end{equation}
with
$\delta$ defined by $\cos \delta = J_0/\sqrt{J_0^2 + D_0^2}$ and $\sin \delta = - D_0/\sqrt{J_0^2 + D_0^2}$.
In this convention, $J_{nm}^\mathrm{eff} > 0$ favors a $\pi$-junction.
The local $\mathrm{U}(1)$ phase at each grain will reorient to satisfy $\cos \Delta \varphi_{nm} = - \mathrm{sgn} (J_{nm}^\mathrm{eff})$ to minimize the total Josephson free energy, such that
\begin{math}
    F_J = - |J^\mathrm{eff}_{1,2}| - |J^\mathrm{eff}_{2,3}| - |J^\mathrm{eff}_{3,1}|.
\end{math}
In an optimal configuration, the $d$ vector orientations between grains $n$ and $m$ satisfy $\beta_{nm} = \pi - \delta$ and do not trap $\pi$-flux; however, the system is limited by the following loop constraint,
\begin{equation}
    \beta_{1,2} + \beta_{2,3} + \beta_{3,1} = 0 \mod 2\pi,
    \label{loop_constraint.three_grain}
\end{equation}
which is typically incompatible with the optimal configuration and consequently leads to a frustrated $d$ vector texture.
The system will trap $\pi$-flux provided that $\sum_{\langle n, m \rangle} \Delta \varphi_{nm} = \pi \mod 2\pi$, yielding the following condition for $\pi$-flux trapping,
\begin{equation}
    \prod_{\langle n, m \rangle}
    \mathrm{sgn}(J^\mathrm{eff}_{nm}) > 0.
    \label{pi-flux_condition}
\end{equation}
In other words, an odd number of $\pi$-junctions induced by the $d$ vector texture can trap half-flux quantum.

Above a critical value of DM-like Josephson coupling, a noncollinear, frustrated $d$ vector texture is preferable at the level of the Josephson free energy and the system can trap $\pi$-flux.
For $J_0 < 0$, the collinear state has $F_\mathrm{col} = -3 |J_0|$, whereas the favored chiral state has free energy $F_\mathrm{chir} = -3 (|J_0|+ \sqrt{3} |D_0|)/2$.
This leads to the critical value $D_* = |J_0|/\sqrt{3}$, at which
the effective Josephson couplings in Eq.~\eqref{effective_josephson_coupling} will change signs for the lowest energy configuration, \textit{i.e.} the emergent $\mathbb{Z}_2$ geometric phase changes its value.
In the regime of $|D_0|< |D_*|$, the $d$ vector texture is ``unfrustrated''---namely, the Heisenberg-like coupling amplitude $J_0$ dominates and promotes collinear $d$ vectors, as shown in Fig.~\ref{fig:d-vector_configurations}(a).
For $J_0 < 0$, it follows that the three-grain superconducting ring is phase-coherent
and does not trap flux for the collinear $d$ vector texture.
In contrast, above the critical value $|D_0| > D_*$, the system is frustrated and energetically prefers a chiral $120^\circ$-ordered texture,
which traps $\pi$-flux for $J_0 < 0$, as shown in Figs.~\ref{fig:d-vector_configurations}(b) and (c).
The chiral $d$ vector texture is analogous to three-sublattice spin systems with antiferromagnetic Heisenberg-like exchange, which similarly feature 120$^\circ$-ordered textures~\cite{Harris1992,Collins1997}.
However, unlike the $120^\circ$-ordered antiferromagnetic configuration in spin systems, a Heisenberg-like coupling is insufficient to realize a chiral $d$ vector texture.
Rather, it is necessary to have, for example, antisymmetric DM-like coupling for the chiral $120^\circ$-ordered $d$ vector texture to be energetically favorable.

More broadly, for an $n$-grain ring, the DM-like coupling prefers a fixed relative twist of the $d$ vectors at each junction, which accumulates around the ring.
The first transition between adjacent sectors with alternating trapped flux occurs at $|D_*/J_0| = \tan (\pi/2n)$~\cite{SM}.
Hence, increasing the number of grains lowers the DM-coupling amplitude required to change the flux sector, applicable to even and odd $n$.
Thus, even a weak local tendency towards noncollinear $d$ vectors can produce spontaneous $\pi$-flux trapping in sufficiently large Josephson networks.

\begin{figure}
    \centering
    \includegraphics[width=\linewidth]{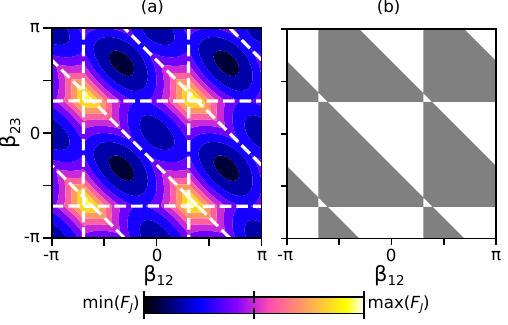}
    \caption[Flux trapping and frustration in a three-grain network]{Flux trapping and frustration in three-grain network of triplet superconductors.
    \textbf{(a)}~Contour plot of the free energy $F_J$ in Eq.~\eqref{free_energy.three_grain.real_D0} as a function of the relative $d$ vector orientation angles $\beta_{1,2}$ and $\beta_{2,3}$ for $D_0 = -0.7 J_0$, which is above the critical value $D_* = |J_0|/\sqrt{3}$.
    \textbf{(b)}~Phase diagram of regions in which $\pi$-flux trapping occurs (gray), satisfying Eq.~\eqref{pi-flux_condition}.
    White dashed lines, corresponding to the boundaries of regions which trap $\pi$-flux, are shown in (a) for comparison.
    The $d$ vector configuration with minimum Josephson free energy for $D_0 > D_*$ and $J_0 < 0$ occurs at $(\beta_{1,2}, \beta_{2,3}) = (2\pi/3, 2\pi/3)$, which traps half-flux quantum, corresponding to the configuration in Fig.~\ref{fig:d-vector_configurations}(b).
    }
    \label{fig:flux_trapping}
\end{figure}

As shown in Fig.~\ref{fig:flux_trapping}(a), the energetic minimum for $D_0 > D_*$ occurs away from the collinear configuration.
For $D_0/J_0<0$, the energetic minimum occurs at the $120^\circ$-ordered $d$ vector texture, with $\beta_{1,2} = \beta_{2,3} = \beta_{3,1} = 2\pi/3$, up to the periodicity of the free energy.
In contrast to the collinear configuration ($\beta_{1,2} = \beta_{2,3} = \beta_{3,1} = 0$) which is not frustrated and does not trap $\pi$-flux, the chiral $120^\circ$-ordered $d$ vector configuration traps half-flux quantum, as shown in Fig.~\ref{fig:flux_trapping}(b).
The $\pi$-flux trapping indicates a spontaneous time-reversal symmetry breaking arising from the frustrated chiral $d$ vector texture.
Moreover, the free energy in Eq.~\eqref{free_energy.three_grain.real_D0} and condition for $\pi$-flux trapping are independent of the sign of the $d$ vector.
For example, under $\beta_1 \rightarrow \beta_1 + \pi$, corresponding to $\hat{\mathbf{d}}_1 \rightarrow - \hat{\mathbf{d}}_1$, it follows that $\beta_{1,2} \rightarrow \beta_{1,2} + \pi$ and $\beta_{3,1} \rightarrow \beta_{3,1} - \pi$.
The $\mathrm{U}(1)$ phase differences will reorient in such a way that Eq.~\eqref{free_energy.three_grain.real_D0} is minimized, and an even number of $\pi$-junctions are introduced, such that the $\pi$-flux condition in Eq.~\eqref{pi-flux_condition} is unchanged.
Consequently, the $d$ vector textures are energetically equivalent up to a $\pi$-rotation.

Above the critical value, the relative sign of $D_0$ determines the chirality of the $d$ vector texture.
For 
$D_0/J_0 < 0$, the minimum energy configuration is a $120^\circ$-ordered coplanar $d$ vector texture, and for $D_0/J_0 > 0$, the minimum energy configuration has opposite vector chirality, as shown in Fig.~\ref{fig:d-vector_configurations}.
The relative sign of the chirality can also be understood from the antisymmetric DM-like coupling $D_0 \hat{\mathbf{z}} \cdot (\hat{\mathbf{d}}_n \times \hat{\mathbf{d}}_m)$, in which the sign of $D_0$ determines the relative orientation of $\hat{\mathbf{d}}_n$ and $\hat{\mathbf{d}}_m$.
Physically, the DM-like interaction favors a fixed rotation between $d$ vectors, while the loop constraint enforces zero total rotation around the triangle, producing geometric frustration.
In the limit $|D_0|\gg |J_0|$, both chiral and antichiral $d$ vector configurations are energetically equivalent, with the $\pi$-flux trapping condition dependent on the sign of $D_0$.

When the system does not have full spin-rotation symmetry, the $d$ vector generally has a preferred orientation determined by, for example, spin-orbit coupling or local spin textures~\cite{Sigrist1991,Gorkov2001,Mackenzie2003,Smidman2017}.
This leads to bulk free-energy terms that penalize deviations from the preferred $d$ vector orientation.
These pinning terms scale with the grain volume and are associated with the condensation energy scale, whereas the Josephson free energy scales with the surface of the grain boundary and is set by the Josephson coupling~\cite{Frazier2026b}.
The proposed planar texture requires sufficiently strong easy-plane confinement and weak residual in-plane pinning compared to Josephson free energy.
The competition between the Josephson free energy and bulk pinning terms can therefore affect the energetics of the frustrated $d$ vector textures, but these pinning terms do not modify the flux trapping conditions.
Flux trapping from frustrated $d$ vector textures is expected to be most favorable when superconducting grains are sufficiently small or the $d$ vector pinning is weak.
More generally, an $n$-grain ring admits distinct textures, and locally pinned textures need not relax to minimize the Josephson free energy.
For any fixed texture $\{\beta_{nm}\}$, the flux-trapping condition is determined by the condition in Eq.~\eqref{pi-flux_condition} for odd $n$.

Lastly, we contrast the $\pi$-flux trapping in the present system with previously studied half-quantum vortices in spin triplet superconductors.
Firstly, both systems are applicable to unitary pairing orders, satisfying $\hat{\mathbf{d}}_n \times \hat{\mathbf{d}}_n^* = 0$.
In the typical case of half-quantum vortices~\cite{Volovik1976,Ivanov2001,Alicea2012}, the $d$ vector does a half-winding about a singular point and is accompanied by a $\pi$-phase winding of the $\mathrm{U}(1)$ phase, such that the pairing order is single-valued.
In contrast, the frustrated $d$ vector in this work has a full $2\pi$-rotation of the $d$ vector, with its chirality determined by the relative signs of the DM-like and Heisenberg-like $d$ vector couplings in Eq.~\eqref{J_nm.three_grain.J0_and_D0}, as seen in Fig.~\ref{fig:d-vector_configurations}.
The $\pi$-flux trapping, corresponding to $\pi$-phase winding of the $\mathrm{U}(1)$ order parameter,
arises purely from frustration of $d$ vectors and is applicable to discrete Josephson junction networks.
Hence, although the present frustrated $d$ vector texture can likewise trap $\pi$-flux for unitary pairing order, the origin is distinct and offers a new platform to realize half-quantum flux trapping in spin triplet superconductors.

 \phantomsection
\paragraph*{Conclusion---}
\addcontentsline{toc}{section}{Conclusion}

We have demonstrated that anisotropic Josephson couplings in networks of spin triplet superconducting grains can lead to frustrated $d$ vector textures that spontaneously produce Josephson currents and can trap nonintegral magnetic flux.
These anisotropic couplings give rise to an emergent geometric phase originating from $d$ vector textures, directly linking the network frustration to the internal pairing structure of the Cooper pair.
For a three-grain network,
above a critical amplitude of DM-like Josephson coupling, the system will energetically prefer a chiral $120^\circ$-ordered $d$ vector texture 
that produces frustrated Josephson phases and spontaneously traps nonintegral flux.
Unlike conventional half-quantum vortices in spin triplet superconductors, the $\pi$-flux trapping originates purely from frustrated $d$ vector textures.

This work introduces a largely unexplored platform to engineer frustrated superconducting systems and $d$ vector textures by controlling anisotropic Josephson couplings and is pertinent to spin triplet superconducting systems in proximity to noncollinear magnetic textures.    
Candidate materials include, for example, superconducting Mn$_3$Ge in proximity to Nb~\cite{Jeon2021, Jeon2023}, as well as $4$Hb-TaS$_2$, consisting of alternating layers of spin-liquid candidate $1T$-TaS$_2$ and Ising superconductor $1H$-TaS$_2$~\cite{Ribak2017, Persky2022, Silber2024, Almoalem2024, Almoalem2024a}.
Both systems are argued to support spin triplet pairing correlations and host noncollinear magnetic textures, which can give rise to anisotropic Josephson couplings and realize frustrated $d$ vector textures.
The proposed flux trapping in this work can be detected using scanning-SQUID microscopy, or by Little-Parks oscillations.

More broadly, the Josephson networks in 
Eq.~\eqref{free_energy.xy_analogy} map to an $XY$ model in the presence of a background gauge field~\cite{Villain1977, Teitel1983}, suggesting the possibility of disorder-induced glassy superconducting phases.
Here, the frustration arises intrinsically from the $d$ vector texture itself rather than being imposed by a fixed background gauge flux or bond phases.
As the coupling 
is sensitive to the grain boundary---including its geometry and impurities---this can give rise to random magnitudes and phase offsets in the Josephson coupling.
Combined with the emergent geometric phase arising from the $d$ vector textures, this platform offers a natural avenue towards realizing frustrated and glassy superconducting states.

\vspace{1.5em}
\phantomsection
\paragraph*{Acknowledgements---}
\addcontentsline{toc}{section}{Acknowledgements}
We thank Oleg Tchernyshyov for his insight on the underlying SEA structure in the role of frustration.
We acknowledge the support of the NSF CAREER Grant No.~DMR-1848349. 
G.R.F. acknowledges 
the hospitality and support of the Kavli Institute for Theoretical Physics,
supported in part by the Heising-Simons Foundation, the Simons Foundation, and grant No. NSF PHY-2309135.

\end{document}


\title{
Supplemental Material for
``Nonintegral
Flux Trapping in Frustrated 
Josephson 
Networks of 
Triplet
Superconductors''
}

\author{Grayson R. Frazier}
\affiliation{Department of Physics and Astronomy, Johns Hopkins University, Baltimore, Maryland 21218, USA}
\affiliation{Kavli Institute for Theoretical Physics, University of California, Santa Barbara, CA 93106, USA}
\author{Colton Lelievre}
\affiliation{Department of Physics and Astronomy, Johns Hopkins University, Baltimore, Maryland 21218, USA}
\author{Yi Li}
\affiliation{Department of Physics and Astronomy, Johns Hopkins University, Baltimore, Maryland 21218, USA}

\maketitle

\onecolumngrid

\setcounter{equation}{0}
\setcounter{section}{0}
\setcounter{figure}{0}
\setcounter{table}{0}
\setcounter{page}{1}
\makeatletter
\renewcommand{\theequation}{S\arabic{equation}}
\renewcommand{\thefigure}{S\arabic{figure}}
\renewcommand{\thesection}{S\arabic{section}}

\section{Validity of Neglecting Magnetic Self-Energy}

In the main text, we neglected the magnetic self-energy associated with the circulating supercurrent.
Here, we establish the regime in which this approximation is justified.
Consider an $n$-junction ring with identical Josephson energies, $E_J = \hbar I_c/ (2e) =  \Phi_0 I_c/(2\pi)$, where $I_c$ is the critical current at each junction.

Let $\Phi_\mathrm{geo}$ denote the preferred flux corresponding to the accumulated geometric phase in the absence of magnetic self-energy, modulo $\Phi_0$,
and let $\epsilon_j$ denote small deviations from the preferred junction phases.
From the single-valuedness loop constraint, it follows that
\begin{equation}
    \sum_{j=1}^n
    \epsilon_j
    =
    \frac{2\pi}{\Phi_0}
    (
    \Phi_\mathrm{geo} - \Phi),
\end{equation}
in which $\Phi$ is the magnetic flux threading the loop.
Expanding the Josephson energy to quadratic order, the Josephson free energy is given by
\begin{equation}
    F_J
    =
    -E_J \sum_{j=1}^n \cos \epsilon_j
    \approx
    -n E_J + 
    \frac{E_J}{2}
    \sum_{j=1}^n \epsilon_j^2.
\end{equation}
For identical junctions, minimizing the quadratic Josephson energy subject to the loop constraint distributes the phase deviations equally, $\epsilon_j = (2\pi/ \Phi_0)(\Phi_\mathrm{geo} - \Phi)/n$.
Consequently, the Josephson free energy is given by
\begin{equation}
    F_J 
    \approx
    -n E_J
    +
    \frac{(\Phi - \Phi_\mathrm{geo})^2}{2nL_J},
\end{equation}
in which
\begin{math}
    L_J
    \equiv
    {\Phi_0}/({2\pi I_c})
\end{math}
is the Josephson inductance of an individual junction.
Accounting for the magnetic self-energy, the free energy is given by
\begin{equation}
    F = F_J + \frac{\Phi^2}{2L}
    \approx
    -nE_J
    +
    \frac{(\Phi - \Phi_\mathrm{geo})^2}{2n L_J}
    +
    \frac{\Phi^2}{2L},
\end{equation}
in which $L$ is the loop self-inductance.

Minimizing the free energy with respect to $\Phi$ yields the following,
\begin{equation}
    \Phi =
    \frac{L}{L + nL_J}
    \Phi_\mathrm{geo}.
\end{equation}
Hence, the trapped flux approaches its preferred value $\Phi_\mathrm{geo}$ in the limit that $L \gg n L_J$, or equivalently,
\begin{equation}
    \frac{2\pi L I_c}{\Phi_0} \gg n.
\end{equation}
For smaller self-inductance, the magnetic flux is suppressed relative to its preferred value.
So far, we have assumed a fixed $d$ vector texture.
More broadly, allowing the texture to relax may additionally modify the ground state and the boundaries at which noncollinear textures can become favorable.

 \section{SEA Structure and the Role of Easy-Plane Anisotropy}
\label{appendix:SEA_comparison}

In this section, we clarify the relationship between the Heisenberg-, Dzyaloshinskii--Moriya (DM)-, and $\Gamma$-type anisotropic Josephson couplings appearing in Eq.~\eqref{dvector_couplings} and the connection to the Shekhtman--Entin-Wohlman--Aharony (SEA) structure.
We further clarify when the latter $\Gamma$-type coupling can vanish in the easy-plane limit.

\subsection{Microscopic Origin of Josephson Couplings}
\label{SM:sec:microscopic_josephson_couplings}

We first provide a summary of the microscopic origin of the Josephson couplings between $d$ vectors,
which largely follows from Refs.~\citenum{Frazier2026} and \citenum{Frazier2026b}.

The electron tunneling between superconducting grains $n$ and $m$ can generically be described by the following tunneling Hamiltonian,
\begin{equation}
    H_{T; nm} = \sum_{\mathbf{k}, \mathbf{k}'}
    \mathbf{c}^\dagger_{n, \mathbf{k}}
    T_{nm}(\mathbf{k}, \mathbf{k}')
    \mathbf{c}_{m, \mathbf{k}'}
    +
    \mathrm{h.c.},
\end{equation}
in which $\mathbf{c}_{n, \mathbf{k}} = (c_{n, \mathbf{k}, \uparrow}, c_{n, \mathbf{k}, \downarrow})^\mathrm{T}$ is the annihilation operator for an electron spinor at grain $n$ with momentum $\mathbf{k}$, and 
\begin{equation}
    T_{nm}(\mathbf{k}, \mathbf{k}')
    =
    T_{nm; 0}(\mathbf{k}, \mathbf{k}') \sigma^0
    +
    \mathbf{T}_{nm}(\mathbf{k}, \mathbf{k}')
    \cdot
    \boldsymbol{\sigma}
\end{equation}
is the single-particle tunneling matrix from grain $n$ to grain $m$.
Here, $T_{nm; 0}$ and $\mathbf{T}_{nm}$ detail spin-independent and spin-dependent tunneling processes, with the latter arising in the presence of, for example, exchange fields or spin-orbit coupling.
From Ambegaokar--Baratoff formalism of Josephson tunneling~\cite{Ambegaokar1963, Sigrist1991, Frazier2024}, it follows that the free energy describing the Josephson coupling between grains $n$ and $m$ is given by
\begin{equation}
    F_{nm} = \mathrm{Re} \left(
    \sum_{\mathbf{k}, \mathbf{k}'}
    \mathcal{J}_{nm}(\mathbf{k}, \mathbf{k}')
    \right).
\end{equation}
Here, the Josephson form factor describing the first-order Josephson coupling is given by
\begin{equation}
    \mathcal{J}_{nm}
    (\mathbf{k}, \mathbf{k}')
    =
    \frac{1}{\beta}
    \sum_{i \omega_n}
    \mathrm{Tr}
    \Big[
    \mathcal{F}_n(\mathbf{k}; i\omega_n)
    [T_{nm}(-\mathbf{k}, -\mathbf{k}')]^\mathrm{T}
    [\mathcal{F}^\dagger_m(\mathbf{k}'; i\omega_n)]^\mathrm{T}
    T_{nm}(\mathbf{k}, \mathbf{k}')
    \Big],
    \label{SMEq:josephson_coupling_GF}
\end{equation}
in which the sum is taken over Matsubara frequencies.
Above, $\mathcal{F}_n(\mathbf{k}, i\omega_n)$ is the anomalous Green's function describing pairing correlations in the particle-particle channel, whose spin-antisymmetric and spin-symmetric components correspond to spin singlet and triplet pairing channels, respectively.

Of relevance to the spin triplet networks discussed in the main text, we consider the Josephson coupling of $d$ vectors, corresponding to the spin-symmetric parts of the anomalous Green's functions.
Generically, the Josephson coupling allows symmetric isotropic, antisymmetric, and symmetric anisotropic contributions,
\begin{equation}
    \mathcal{J}_{nm}
    (\mathbf{k}, \mathbf{k}')
    =
    J_{nm}(\mathbf{k}, \mathbf{k}')
    \mathbf{d}_n(\mathbf{k})
    \cdot
    \mathbf{d}_m^*(\mathbf{k}')
    +
    \mathbf{D}_{nm}(\mathbf{k}, \mathbf{k}')
    \cdot
    (\mathbf{d}_n(\mathbf{k}) \times \mathbf{d}_m^*(\mathbf{k}')
    )
    +
    \mathbf{d}_n^\mathrm{T}(\mathbf{k})
    \Gamma_{nm}(\mathbf{k}, \mathbf{k}')
    \mathbf{d}_m^*(\mathbf{k}').
\end{equation}
The three couplings can be identified with Heisenberg-like, DM-like, and $\Gamma$-type couplings of the $d$ vector, respectively.
In the case that the spin and momentum-dependence can be decoupled, the Josephson coupling coefficients can be decomposed as follows,
\begin{equation}
    J_{nm} 
    \propto 
    T_{nm;0}^2 + \mathbf{T}_{nm} \cdot \mathbf{T}_{nm},
    \hspace{1.5em}
    \mathbf{D}_{nm}
    \propto
    i T_{nm; 0} \mathbf{T}_{nm},
    \hspace{1.5em}
    \Gamma_{nm}
    \propto
    \mathbf{T}_{nm} \mathbf{T}_{nm}^\mathrm{T},
    \label{SMEq:josephson_couplings_T0_T_parameterization}
\end{equation}
following from Eq.~\eqref{SMEq:josephson_coupling_GF}.
Hence, for \emph{single-channel} tunneling, where all Josephson couplings are determined by the same tunneling matrix, the three couplings are not independent.
Rather, they correspond to an underlying rotation of the $d$ vector, as we discuss in the following section.

\subsection{SEA Structure}

For unitary $d$ vectors, the bilinear Josephson coupling can be written generally as
\begin{equation}
    \mathcal{J}_{nm} = e^{i \Delta \varphi_{nm}} \hat{\mathbf{d}}_n^\mathrm{T} K_{nm} \hat{\mathbf{d}}_m,
\end{equation}
as discussed in the main text.
Here, the coupling matrix can generally be decomposed into isotropic, antisymmetric, and symmetric anisotropic channels,
\begin{equation}
    K_{nm}^{ab} = J_{nm} \delta_{ab} + \epsilon_{abc} D_{nm}^c + \Gamma_{nm}^{ab}.
\end{equation}

For a rotation-locked junction, these three terms are not independent.
Specifically, they can be parameterized as
\begin{equation}
    J_{nm} = c(w^2 - \mathbf{v}^2),
    \hspace{1.5em}
    \mathbf{D}_{nm} = 2 cw \mathbf{v},
    \hspace{1.5em}
    \Gamma_{nm}^{ab} = 2 c v_a v_b.
\end{equation}
This construction is analogous to that derived from Ambegaokar--Baratoff formalism, as shown in Sec.~\ref{SM:sec:microscopic_josephson_couplings},
with $w$ and $\mathbf{v}$ playing the role of the spin-independent and spin-dependent single-particle tunneling, respectively.
For example, the Josephson couplings in Eq.~\eqref{SMEq:josephson_couplings_T0_T_parameterization} obey this parameterization:
identifying $w \propto T_{nm;0}$ and $\mathbf{v} \propto i\mathbf{T}_{nm}$, the symmetric anisotropic coupling satisfies $\Gamma_{nm} \propto \mathbf{D}_{nm}\mathbf{D}_{nm}^\mathrm{T}$.
In general $w$ and $\mathbf{v}$ are complex, carrying both the magnitude and phase accumulated in the microscopic tunneling process.
The full coupling matrix is therefore given by
\begin{equation}
    K_{nm} = c R(w, \mathbf{v}),
\end{equation}
with
\begin{equation}
    [R(w, \mathbf{v})]_{ab}
    =
    (w^2 - \mathbf{v}^2) \delta_{ab} + 2 v_a v_b + 2 w \epsilon_{abc} v_c.
\end{equation}
For real $w$ and $\mathbf{v}$, this is proportional to an $\mathrm{SO}(3)$ matrix, satisfying $R^\mathrm{T} R = (w^2 + \mathbf{v}^2)^2 \mathds{1}$, with $R/(w^2 + \mathbf{v}^2) \in \mathrm{SO}(3)$.
Thus, in the rotation-locked limit, the Heisenberg-, DM-, and $\Gamma$-type couplings together describe a single rotated coupling of the $d$ vectors.

The structure presented here is directly analogous to the hidden symmetry identified by Shekhtman, Entin-Wohlman, and Aharony in the context of magnetic superexchange~\cite{Shekhtman1992}.
In that problem, spin-orbit coupling produces a DM interaction, together with a related symmetric anisotropic exchange~\cite{Dzyaloshinsky1958,Moriya1960}.
When their coefficients satisfy the appropriate relation, the full Hamiltonian describing the coupling at the bond can be transformed by local rotations of the two spins into an isotropic gauged Heisenberg interaction.
In other words, the coupling
\begin{math}
    \mathcal{H}_{ij}
    =
    J \mathbf{S}_i \cdot \mathbf{S}_j
    +
    \mathbf{D} \cdot (\mathbf{S}_i \times \mathbf{S}_j)
    +
    \mathbf{S}_i^\mathrm{T} \Gamma \mathbf{S}_j
\end{math}
can be rewritten as
\begin{equation}
    \mathcal{H}_{ij}
    =
    \tilde{J} \mathbf{S}_i \cdot R_{ij} \mathbf{S}_j.
\end{equation}
The same hidden-rotation structure appears here for unitary $d$ vectors and real-valued couplings,
\begin{equation}
    J \hat{\mathbf{d}}_n \cdot \hat{\mathbf{d}}_m
    +
    \mathbf{D} \cdot 
    \left(
    \hat{\mathbf{d}}_n \times \hat{\mathbf{d}}_m
    \right)
    +
    \hat{\mathbf{d}}_n^\mathrm{T} \Gamma \hat{\mathbf{d}}_m
    =
    \tilde{J} \hat{\mathbf{d}}_n \cdot R_{nm} \hat{\mathbf{d}}_m.
    \label{Josephson_coupling.rotation}
\end{equation}
The apparent anisotropic couplings therefore represent, in this limit, different components of a single rotated $d$ vector interaction rather than independent anisotropies.
This is closely related to the gauged Heisenberg models in spin-orbit coupled spin systems~\cite{Hill2021}.

The symmetric anisotropic $\Gamma$-type coupling has a particularly simple form.
Given that $\Gamma_{nm} = 2 c \mathbf{v} \mathbf{v}^\mathrm{T}$ and $\mathbf{D}_{nm} = 2cw \mathbf{v}$, it follows that
$\Gamma_{nm} \propto \mathbf{D}_{nm} \mathbf{D}_{nm}^\mathrm{T}$.
Consequently, if an easy-plane anisotropy confines the $d$ vectors to the plane normal to the DM vector, then the $\Gamma$-type coupling vanishes, $\hat{\mathbf{d}}_n^\mathrm{T} \Gamma_{nm} \hat{\mathbf{d}}_m \propto (\hat{\mathbf{d}}_n \cdot \mathbf{D}_{nm})(\hat{\mathbf{d}}_m \cdot \mathbf{D}_{nm}) = 0$.
Therefore, easy-plane confinement reduces the couplings to solely Heisenberg- and DM-like terms.
Hence, disregarding the $\Gamma$-type coupling is exact in the easy-plane limit, rather than being a perturbative result.

The role of the easy-plane confinement can also be seen clearly as follows.
If the real unitary $d$ vectors have full $\mathrm{SO}(3)$ symmetry, then they may freely rotate to satisfy the local Josephson coupling in Eq.~\eqref{Josephson_coupling.rotation}.
Namely, a configuration that optimizes the local coupling satisfies $\hat{\mathbf{d}}_n = R_{nm} \hat{\mathbf{d}}_m$, up to a $\mathbb{Z}_2$ redundancy from the $d$ vector.
Transporting the $d$ vector around a closed ring produces the holonomy~\cite{Wilson1974a, Toulouse1977},
\begin{equation}
    W = \prod_{\langle n m \rangle} R_{nm}.
\end{equation}
Single-valuedness requires $W \hat{\mathbf{d}}_1 \propto \hat{\mathbf{d}}_1$, with any sign differences absorbed by the $\mathrm{U}(1)$ phase.
Every $\mathrm{SO}(3)$ matrix leaves an axis invariant;
hence, the texture can satisfy $W \hat{\mathbf{d}}_1 = \hat{\mathbf{d}}_1$ by aligning with the rotation axis of $W$, optimizing every rotation-locked bond simultaneously and removing the frustration.

This formulation provides an extension of the SEA picture to a closed Josephson network.
At a single Josephson junction, the hidden SEA symmetry allows the DM and $\Gamma$-type terms to be absorbed into a relative rotation of local frames.
Around a loop, these rotations need not compose to identity but rather can generate a nontrivial holonomy.
However, for a $d$ vector with full $\mathrm{SO}(3)$ symmetry, there is an out-of-plane escape that prevents bond rotations alone from producing frustration.
An easy-plane anisotropy precisely removes the out-of-plane escape.
Taking $\mathbf{D}_{nm} = D_0 \hat{\mathbf{z}}$ and restricting the $d$ vector to the $xy$ plane, the $\Gamma$-type coupling vanishes exactly.
The out-of-plane axis is excluded from the possible $d$ vector configurations.
The planar frustrated $d$ vector system studied in the main text is recovered exactly.
More broadly,
when unitary $d$ vectors have full $\mathrm{SO}(3)$ orientational freedom, the rotation-locked couplings alone do not frustrate the $d$ vector texture for a closed loop.
Rather, any frustration must originate from additional constraints, or from signs of the Josephson couplings.

This analogy holds for real-valued $w$ and $\mathbf{v}$, to which we have restricted ourselves above.
In this rotation-locked case, an overall complex phase factors out, and the remaining coupling matrix is proportional to a real $\mathrm{SO}(3)$ rotation.
More generally $w$ and $\mathbf{v}$, inherited from the microscopic tunneling amplitudes, are complex, as shown in Refs.~\citenum{Frazier2026} and \citenum{Frazier2026b}.
The relation is then $R(w, \mathbf{v})/(w^2 + \mathbf{v}^2) \in \mathrm{SO}(3, \mathbb{C})$, which still preserves the bilinear $\hat{\mathbf{d}}_n^\mathrm{T} \hat{\mathbf{d}}_m$.

 \section{Relaxation of \texorpdfstring{$d$}{d} Vectors in the Presence of Complex Coupling Amplitudes}
\label{appendix:nonintegral_flux_relaxed_d_vectors}

In the main text, the nonintegral phase accumulated in a loop was first considered for a fixed $d$ vector orientation.
Here, we provide an example in which the nonintegral flux trapping survives when $d$ vector orientations are themselves allowed to relax, for complex-valued Josephson coupling amplitudes.
We focus on the case of unitary $d$ vectors in presence of purely imaginary DM-like Josephson coupling.

Consider three identical junctions in a ring geometry, with Josephson couplings $J_0 < 0$ and $\mathbf{D}_{nm} = i \tilde{D}_{0} \hat{\mathbf{z}}$, where $\tilde{D}_0 \in \mathbb{R}$.
We assume an easy-plane anisotropy that confines the real unitary $d$ vectors to the $xy$ plane,
writing $\hat{\mathbf{d}}_n = (\cos \beta_n, \sin \beta_n, 0)^\mathrm{T}$ and $\beta_{nm} \equiv \beta_m - \beta_n$.
For a symmetric anisotropic coupling with $\Gamma_{nm} \propto \hat{\mathbf{z}} \hat{\mathbf{z}}^\mathrm{T}$, the $\Gamma$-type matrix element vanishes in the easy-plane limit, and the phase-independent Josephson form factor is given by
\begin{equation}
    \tilde{\mathcal{J}}_{nm} = J_0 \cos \beta_{nm} + i \tilde{D}_0 \sin \beta_{nm},
\end{equation}
subject to the constraint $\beta_{12} + \beta_{23} + \beta_{31} = 0 \mod 2\pi$.

Minimization over the superconducting phase differences reduces the Josephson free energy to $F_J = - \sum_{\langle n, m \rangle} |\tilde{\mathcal{J}}_{nm}|$, in which
\begin{equation}
    |\tilde{\mathcal{J}}_{nm}|^2
    =
    J_0^2 \cos^2 \beta_{nm} + \tilde{D}_0^2 \sin^2 \beta_{nm}
    =
    J_0^2 + (\tilde{D}_0^2 - J_0^2) \sin^2 \beta_{nm}.
    \label{SMeq:form_factor_magnitude.imaginary_D}
\end{equation}
In contrast to the real-valued couplings discussed in Sec.~\ref{appendix:critical_josephson_coupling}, the magnitude depends on $\beta_{nm}$ only through $\sin^2 \beta_{nm}$ and is therefore even in $\beta_{nm}$.
For $|\tilde{D}_0| < |J_0|$, the coefficient of $\sin^2 \beta_{nm}$ is negative, and each Josephson coupling has an upper bound $|J_0|$, and a collinear $d$ vector texture optimizes every bond simultaneously.
This is the case realized by a single rotation-locked tunneling channel, for which $J_{nm}$ and $\mathbf{D}_{nm}$ are not independent.
For example, the parameterization discussed in Sec.~\ref{appendix:SEA_comparison} gives $J_{nm} = c(w^2 + |\mathbf{v}|^2)$ and $|\mathbf{D}_{nm}| = 2 c |w| |\mathbf{v}|$ for imaginary $\mathbf{v}$, so that $|J_{nm}| \geq |\mathbf{D}_{nm}|$.

Instead, suppose that $\tilde{D}_0$ and $J_0$ can be tuned independently in this toy model, such that the regime $|\tilde{D}_0| > |J_0|$ can be accessed, and collinear texture may not correspond to an energetic minimum.
The coefficient of $\sin^2 \beta_{nm}$ in Eq.~\eqref{SMeq:form_factor_magnitude.imaginary_D} is then positive, and each bond is optimized at $\beta_{nm} = \pm \pi/2$, which the loop constraint cannot satisfy at all three junctions at once.
To obtain the global minimum, we find a bound on $\sum_{\langle n, m \rangle} \sin^2 \beta_{nm}$, subject to the loop constraint.
Firstly, we use $\sin^2 \beta_{nm} = (1 - \cos 2\beta_{nm})/2$ and hence find a bound on $\sum_{\langle n, m\rangle} \cos (2 \beta_{nm})$.
Writing $x = 2 \beta_{12}$ and $y = 2 \beta_{23}$, the constraint gives $2 \beta_{31} = -(x+y)$.
Defining $t \equiv |\cos \frac{x+y}{2}|$, this yields
\begin{equation}
    \sum_{\langle n, m \rangle} \cos (2 \beta_{nm})
    =
    \cos x + \cos y + \cos (x+y)
    \geq
    -2t + (2t^2 - 1)
    =
    2 \left( t - \frac{1}{2} \right)^2 
    - 
    \frac{3}{2}
    \geq
    - \frac{3}{2}.
    \label{SMeq:cos_bound}
\end{equation}
The first inequality uses $\cos x + \cos y = 2 \cos \frac{x+y}{2} \cos \frac{x-y}{2} \geq -2t$, together with $\cos (x+y) = 2t^2 - 1$.
Since $\sum_{\langle n, m \rangle} \sin^2 \beta_{nm} = \frac{3}{2} - \frac{1}{2} \sum_{\langle n, m \rangle} \cos (2 \beta_{nm})$, it follows that $\sum_{\langle n, m \rangle} \sin^2 \beta_{nm} \leq 9/4$.
Consequently, for $|\tilde{D}_0| > |J_0|$,
the energy bound is given by
\begin{equation}
    \left(
    \sum_{\langle n, m \rangle} |\tilde{\mathcal{J}}_{nm}|
    \right)^2
    \leq
    3 \sum_{\langle n, m \rangle} |\tilde{\mathcal{J}}_{nm}|^2
    \leq
    \frac{9}{4} \left( J_0^2 + 3 \tilde{D}_0^2 \right),
    \label{SMeq:energy_bound}
\end{equation}
with the first inequality being the Cauchy--Schwarz inequality for a three-component vector.
Both inequalities are saturated at $\beta_{nm} = \pm 2\pi/3$,
in which the Josephson couplings at each junction are equal, and $\sum_{\langle n, m \rangle} \sin^2 \beta_{nm} = 9/4$ attains the bound of Eq.~\eqref{SMeq:cos_bound}.
The chiral textures are therefore global minima of the Josephson free energy for $|\tilde{D}_0| > |J_0|$, with
\begin{equation}
    F_J = - \frac{3}{2} \sqrt{J_0^2 + 3 \tilde{D}_0^2}.
    \label{SMeq:free_energy.chiral_imaginary_D}
\end{equation}

For either chirality, the flux trapped in the three-grain loop follows from Eq.~\eqref{gauge_field.flux_condition},
\begin{equation}
    \frac{\Phi}{\Phi_0}
    =
    \pm
    \left[
    \frac{3}{2\pi}
    \arctan
    \left(
    \frac{\sqrt{3} |\tilde{D}_0|}{|J_0|}
    \right)
    - \frac{1}{2}
    \right].
    \label{SMEq:flux_trapped_imaginary_D0}
\end{equation}
The two signs correspond to the two chiral textures, which are energetically degenerate.
Expanding Eq.~\eqref{SMEq:flux_trapped_imaginary_D0} for large $\tilde{D}_0$, we find,
\begin{equation}
    \frac{\Phi}{\Phi_0}
    \approx
    \pm\frac{1}{4}
    \mp
    \frac{\sqrt{3}}{2\pi} \left| \frac{J_0}{\tilde{D}_0} \right|
    +
    \mathcal{O} \left( \left| \frac{J_0}{\tilde{D}_0} \right|^3\right).
\end{equation}
Hence, the trapped flux continuously approaches $\pm \Phi_0/4$ for $|\tilde{D}_0| > |J_0|$.

\begin{figure}
    \centering
    \includegraphics[width=0.5\linewidth]{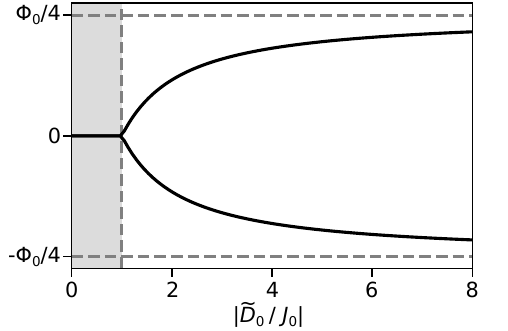}
    \caption{
    Flux trapped for complex DM-like Josephson coupling and real-valued Heisenberg-like Josephson coupling, for easy-plane unitary $d$ vectors.
    In the single-channel rotation-locked limit, with real $w$ and purely imaginary $\mathbf{v}$, the couplings obey $|\tilde{D}_0/J_0| \leq 1$, resulting in a collinear texture with zero trapped flux (gray).
    In contrast, for $|\tilde{D}_0/J_0| > 1$, a chiral texture is favored, which traps nonintegral flux.
    The trapped flux approaches $\pm \Phi_0/4$, with the two chiral textures being energetically degenerate at the level of the Josephson free energy.
    }
    \label{fig:complex_d_flux_trapped}
\end{figure}

As shown in Fig.~\ref{fig:complex_d_flux_trapped},
the trapped flux increases continuously from zero at $|\tilde{D}_0| = |J_0|$ and saturates at $\pm \Phi_0/4$ in the limit $|\tilde{D}_0|/|J_0| \rightarrow \infty$, 
corresponding to a quarter flux quantum trapped.
At the marginal value $|\tilde{D}_0| = |J_0|$, the bond amplitude in Eq.~\eqref{SMeq:form_factor_magnitude.imaginary_D} is independent of $\beta_{nm}$.
At this point, the geometric phase reduces to $\mathcal{A}_{nm} = \mp \beta_{nm}$, so the loop constraint returns integer flux.
Hence, for $|\tilde{D}_0| > |J_0|$, the network traps nonintegral flux even when the $d$ vector orientations are free to relax.

This phenomenological example is a simple toy model demonstrating that relaxation of the $d$ vectors does not generically forbid nonintegral flux trapping.
However, realizing the above couplings beyond the SEA structure, or more broadly for complex couplings, even with a single tunneling channel, remains an open question.

 \section{Critical Josephson Coupling}
\label{appendix:critical_josephson_coupling}

We solve for a critical value of $D_0$ in Eq.~\eqref{free_energy.three_grain.real_D0} for when it becomes energetically favorable for the $d$ vectors to form a noncollinear, frustrated texture.
At this critical value, an odd number of the effective Josephson couplings will change sign.
The total Josephson free energy is given by
\begin{equation}
    F_J = 
    \sum_{\langle n m\rangle}
    J^\mathrm{eff}_{n, m} \cos(\Delta \varphi_{nm}),
    \label{SMEq:free_energy_real_D0_J0}
\end{equation}
in which the effective coupling amplitude is
\begin{equation}
    J^\mathrm{eff}_{n, m} = \sqrt{J_0^2 + D_0^2} \cos (\beta_{nm} + \delta),
    \label{SMeq:effective_josephson_coupling}
\end{equation}
with $\beta_{nm} = \beta_m - \beta_n$ and 
$\delta$ defined by $\cos \delta = J_0/\sqrt{J_0^2 + D_0^2}$ and $\sin \delta = - D_0/\sqrt{J_0^2 + D_0^2}$.

\subsection{\texorpdfstring{$n$}{n}-grain Ring}
\label{appendix:n_grain_ring}

Let us consider a ring of $n$ superconducting grains.
Minimizing Eq.~\eqref{SMEq:free_energy_real_D0_J0} over the $\mathrm{U}(1)$ phase differences yields
\begin{equation}
    F_n
    =
    - \sqrt{J_0^2 + D_0^2}
    \sum_{\langle n m \rangle}
    \left| \cos x_{nm} \right|,
    \hspace{1.5em}
    x_{nm} \equiv \beta_{nm} + \delta,
\end{equation}
which is subject to the loop constraint $\sum_{\langle n m \rangle} x_{nm} = n \delta \mod 2\pi$, which follows from $\sum_{\langle n m \rangle}\beta_{nm} = 0 \mod 2\pi$.
The free energy at each Josephson junction depends only on $|\cos x_{nm}|$, which has period $\pi$ rather than $2\pi$.
This is precisely the $\mathbb{Z}_2$ identification for a unitary $d$ vector.
Namely, a $\pi$ rotation of the $d$ vector changes the sign of the effective coupling on each adjacent coupling, and the $\mathrm{U}(1)$ phase difference absorbs that sign as a $\pi$ phase.

In solving for the energetic minima, we consider all possible $d$ vector configurations, including but not limited to the uniform and symmetric chiral textures.
For simplicity in notation, we write $x_{nm} = y_{nm} + \pi \ell_{nm}$, in which $y_{nm} \in (-\pi/2, \pi/2]$ and $\ell_{nm} \in \{0, 1\}$.
The free energy depends only on the $y_{nm}$ and is insensitive to $\ell_{nm}$, and reducing the loop constraint modulo $\pi$ removes them entirely,
\begin{equation}
    \sum_{\langle n m \rangle} y_{nm} = n \delta \mod \pi.
\end{equation}
Nonetheless, the residual condition modulo $2\pi$ survives and fixes the parity of $\sum_{\langle n m \rangle} \ell_{nm}$, or, equivalently, the parity of the number $n_\pi = \sum_{\langle n m \rangle} \ell_{nm}$ of $\pi$-junctions around the ring.

Given that $\cos y_{nm}$ is concave on the interval $(-\pi/2, \pi/2)$, 
the sum $\sum_{\langle n m \rangle} \cos y_{nm}$ is maximized when all $y_{nm}$ are equal.
The minimal free energy for the $n$ grain ring reduces to
\begin{equation}
    F_n = - n \sqrt{J_0^2 + D_0^2} \cos d,
\label{SMeq:n_grain_closed_form}
\end{equation}
Here, $d = \mathrm{dist}(\delta, \frac{\pi}{n} \mathbb{Z})$ denotes the distance from $\delta$ to the nearest integer multiple of $\pi/n$.
Because $d$ is invariant under $\delta \rightarrow \delta + \pi$, Eq.~\eqref{SMeq:n_grain_closed_form} may equivalently be evaluated with $\delta$ replaced by $\arctan|D_0/J_0| \in [0, \pi/2)$.
This invariance applies to the free energy, but the trapped flux discussed below is not invariant under $\delta \rightarrow \delta + \pi$ for odd $n$.
The texture is collinear when the nearest multiple of $\pi/n$ is itself a multiple of $\pi$, and chiral otherwise.
The two first become degenerate precisely at the critical value $|\delta_*| = \pi/2n$.
At this point, the critical value of the antisymmetric DM-like coupling is given by
\begin{equation}
    D_* = |J_0| \tan \frac{\pi}{2n},
\end{equation}
which decreases with the number of grains, $n$.

More intuitively, the DM-like term prefers a twist of $-\delta \mod \pi$ at each junction, leading to a total twist $-n \delta$.
Equivalently, in terms of $x_{nm} = \beta_{nm} + \delta$, each bond is optimized at $x_{nm} = 0 \mod \pi$, while the loop constraint forces $\sum_{\langle n m \rangle} x_{nm} = n \delta$.
However, the loop constraint only allows the total twisting to be an integral multiple of $\pi$, corresponding to the modulo $\pi$ constraint.
The remaining $n\delta - \pi p$ is distributed equally over all $n$ junctions, in which $p$ is the closest integer to $n\delta/\pi$, giving 
\begin{equation}
    y_{nm} = \delta - \frac{\pi p}{n},
    \hspace{1.5em}
    d = |y_{nm}|.
\end{equation}
The crossing in the free energy in Eq.~\eqref{SMeq:n_grain_closed_form}---corresponding to when another value of $p$ is taken---is determined by
\begin{equation}
    \cos
    \left(
    \delta - \frac{\pi p_*}{n}
    \right)
    =
    \cos
    \left(
    \delta - \frac{\pi (p_*+1)}{n}
    \right),
\end{equation}
for critical value $p_*$.
This determines a critical value of $\delta$ to be
\begin{equation}
    \delta_* = \frac{\pi p_*}{n} + \frac{\pi}{2n},
    \label{SMeq:n_grain_crossings}
\end{equation}
in which $p_* \in \mathbb{Z}$.
Since the effective coupling in Eq.~\eqref{SMeq:effective_josephson_coupling} depends on $\delta_*$ through $|D_0/J_0| = |\tan \delta_*|$, only the $p_*$ with $\delta_* \in [0, \pi/2)$ are obtained by increasing $|D_0|$ relative to $|J_0|$.
The first crossing at $p_* = 0$ separates the collinear $p = 0$ and noncollinear $p=1$  sector.

For example, for $n=3$, the only critical coupling in this range is $\delta_* = \pi/6$, which corresponds to the transition to the chiral configuration discussed in Sec.~\ref{SM:subsec:three_grain_network}.
The next solution, $\delta_* = \pi/2$, sits at the endpoint of the interval, corresponding to $J_0 = 0$, where the two chiral textures are degenerate.
Given $\tan(\pi/6) = 1/\sqrt{3}$, this recovers the critical coupling discussed in the main text.
For $n=4$, this yields critical values $\delta_*$ of $\pi/8$ and $3\pi/8$ in the relevant range, corresponding to values of $|D_0/J_0|$ of $ \sqrt{2} - 1$ and $\sqrt{2} + 1$.
Hence, the $n=4$ grain system traps half-flux quantum in the window $\sqrt{2} - 1 < |D_0/J_0| < \sqrt{2} + 1$.

More generally, the $\pi$-flux trapping condition follows from the same construction.
At the optimum every junction carries $|J^\mathrm{eff}_{nm}| = \sqrt{J_0^2 + D_0^2} \cos d$ with $\mathrm{sgn}(J^\mathrm{eff}_{nm}) = (-1)^{\ell_{nm}}$, such that $\Delta \varphi_{nm} = \pi (1 - \ell_{nm})$.
Hence, the trapped flux is given by,
\begin{equation}
    \frac{\Phi}{\Phi_0}
    =
    - \frac{1}{2}
    \left(
    n - \sum_{\langle n m \rangle} \ell_{nm}
    \right)
    \mod 1.
\end{equation}
The ring therefore traps half-flux quantum when $n - \sum_{\langle nm \rangle} \ell_{nm}$ is odd, and the parity of $\sum_{\langle nm \rangle} \ell_{nm}$ is fixed by the loop constraint.
Specifically, $\sum_{\langle n m \rangle} \ell_{nm} \equiv p \mod 2$, so that half-flux quantum is trapped when $n - p$ is odd.
For $n \geq 4$, the sequence in Eq.~\eqref{SMeq:n_grain_crossings} has more than one solution in this range.
Hence, the trapped flux alternates between successive intervals rather than persisting above a single threshold.
Rings with an even number of grains are therefore not excluded.
As discussed, a four-grain ring traps $\pi$-flux for $\tan(\pi/8) < |D_0/J_0| < \tan(3\pi/8)$, and a six-grain ring for $\tan(\pi/12) < |D_0/J_0| < \tan(\pi/4)$ and again for $|D_0/J_0| > \tan(5\pi/12)$.

An important consequence of the above result is that increasing the number of grains lowers the DM-like coupling required for the onset of half-flux trapping.
In particular, the first transition occurs at
\begin{equation}
    \left |\frac{D_*}{J_0} \right|
    = \tan \left( \frac{\pi}{2n} \right),
    \label{critical_coupling_n_grain}
\end{equation}
which approaches $\pi/2n$ for large $n$.
Thus, even a weak preferred twist at each individual junction can accumulate around a sufficiently large ring and drive the system into a ground state that traps $\pi$-flux.
Physically, the DM-like coupling favors a small relative rotation of $d$ vectors at each junction, and as $n$ increases, a smaller twist per junction is therefore sufficient to cross between adjacent sectors.
Beyond the first transition, the two sectors of alternating flux parity alternate as $|D_0/J_0|$ is increased.

\subsection{Three-Grain Network}

\label{SM:subsec:three_grain_network}

Let us consider the simplest loop, a three-grain network.
In this case, the critical coupling in Eq.~\eqref{critical_coupling_n_grain} reduces to
\begin{equation}
    \left|
    \frac{D_*}{J_0}
    \right|
    =
    \frac{1}{\sqrt{3}}.
\end{equation}
At this critical coupling, the collinear and $120^\circ$-ordered textures become energetically degenerate, with the latter trapping $\pi$-flux.
Below, we consider a few of the simplest cases for illustration.

\begin{figure}
    \centering
    \includegraphics[width=\linewidth]{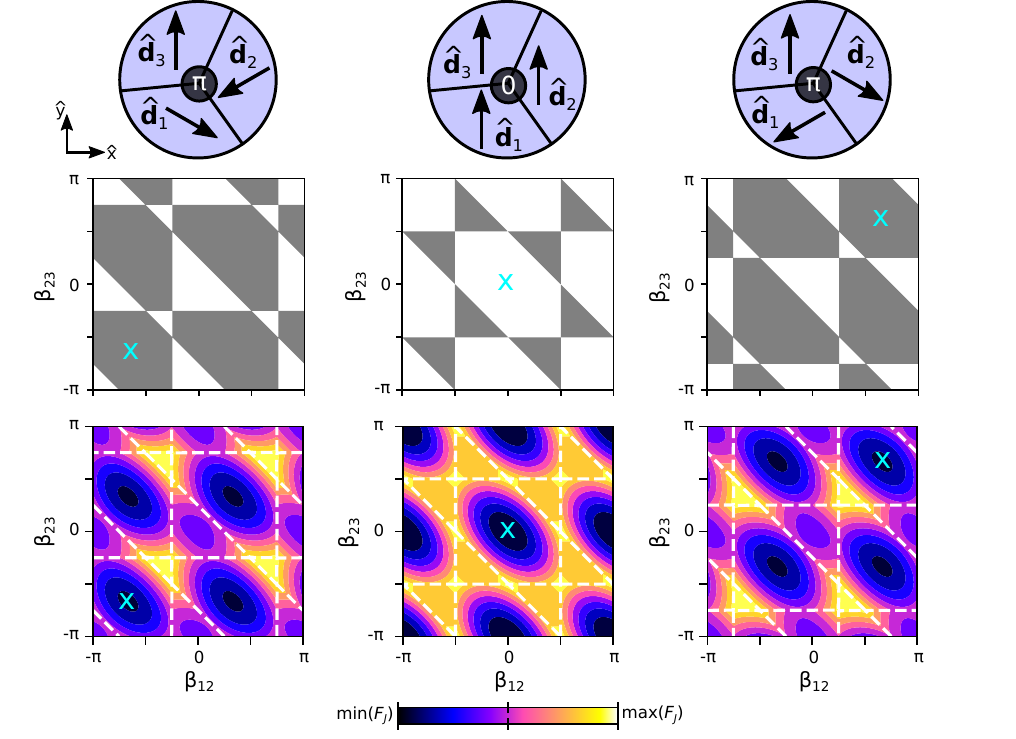}
    \caption[Phase diagrams for representative values of the DM-like coupling $D_0$]{Phase diagrams as a function of relative $d$ vector orientations for representative values of DM-like coupling $D_0$ in Eq.~\eqref{SMEq:free_energy_real_D0_J0}.
    The top row depicts a minimal energy $d$ vector configuration with the trapped flux in black, the middle row includes the shaded regions which trap $\pi$-flux in gray, and the bottom row is a contour plot of the free energy in Eq.~\eqref{SMEq:free_energy_real_D0_J0}.
    The coloring is the same as that of Fig.~\ref{fig:flux_trapping} in the main text.
    \textbf{(Left column)}~The value of DM-like coupling is $D_0 = J_0 < 0$.
    Up to the periodicity of the free energy, the minimal energy $d$ vector configuration (cyan ``x'') corresponds to the antichiral configuration in the top row.
    \textbf{(Middle column)}~Phase diagram for $D_0 = 0$ and  $J_0 < 0$.
    The minimal energy $d$ vector configuration (cyan ``x'') corresponds to the collinear configuration in the top row.
    \textbf{(Right column)}~Phase diagram for $D_0 = -J_0 >0$.
    The minimal energy $d$ vector configuration (cyan ``x'') corresponds to the $120^\circ$-ordered chiral configuration in the top row.
    }
    \label{SMfig:phase_diagram}
\end{figure}

In the regime that $|D_0| < D_*$, it follows that the Heisenberg-like coupling dominates, favoring collinear $d$ vector textures energetically, as shown in the middle column of Fig.~\ref{SMfig:phase_diagram}.
For $D_0 = 0$ and $J_0 < 0$, the free energy is minimized for parallel $d$ vectors, yielding a homogeneous and phase-coherent pairing order with $\beta_{1,2} = \beta_{2,3} = \beta_{3,1} = 0$.
In the case that $J_0 > 0$, it follows that the system will resolve the frustration by adopting an odd number of $\pi$-junctions, leading to $\pi$-flux trapping with (anti)collinear $d$ vectors, $\beta_{1,2} = \beta_{2,3} = \beta_{3,1} = 0$ up to the periodicity of the free energy.

As one increases the strength of $D_0$ relative to $J_0$, local energetic minima will begin to appear away from the collinear configuration ($\beta_{1,2} = \beta_{2,3} = \beta_{3,1} = 0$).
At the critical point, $D_* = \pm J_0/\sqrt{3}$, the free energy for $\beta_{1,2} = \beta_{2,3} = \beta_{3,1} = 0$ is equivalent to that with $\beta_{1,2} = \beta_{2,3} = \pm2\pi/3$.
The relative value of $D_0$ and $J_0$ in this regime determine the chirality of the $d$ vector textures.
For $D_0/J_0> 0$ and $|D_0| > D_*$, an antichiral $120^\circ$-ordered $d$ vector texture which traps $\pi$-flux corresponds to the energetic minimum, as shown in the left column in Fig.~\ref{SMfig:phase_diagram}.
For $D_0/J_0 < 0$ and $|D_0| > D_*$, the system prefers energetically the $120^\circ$-ordered $d$ vector texture, which likewise traps $\pi$-flux, as shown in the right column of Fig.~\ref{SMfig:phase_diagram}.

\begin{figure}
    \centering
    \includegraphics[width=.5\linewidth]{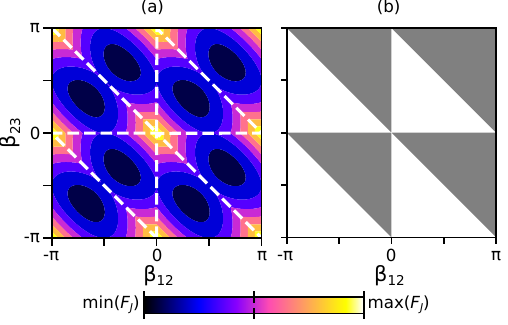}
    \caption[Free energy and $\pi$-flux trapping for $|D_0| \gg |J_0|$]{\textbf{(a)}~Free energy and \textbf{(b)}~$\pi$-flux trapping in the limit of $|D_0| \gg |J_0|$, with $D_0 > 0$.
    The color scheme is the same as Fig.~\ref{fig:flux_trapping} in the main text.}
    \label{SMfig:large_D0}
\end{figure}

Lastly, in the regime that $|D_0| \gg |J_0|$, both chiral $d$ vector configurations correspond to energetic minima, as seen in Fig.~\ref{SMfig:large_D0}.
The sign of $D_0$ determines which chiral configuration can trap $\pi$-flux.
In the case of $J_0=0$ and $D_0 > 0$, the $120^\circ$-ordered chiral configuration (Fig.~\ref{SMfig:phase_diagram}, right column) traps $\pi$-flux, whereas the antichiral configuration (Fig.~\ref{SMfig:phase_diagram}, left column) does not.
The opposite is true for $D_0 < 0$.
Moreover, the amount of regions in the phase diagram which trap $\pi$-flux is enhanced as $D_0$ increases.
For example, in the case that $D_0 \rightarrow 0$, only $25\%$ of the phase space corresponds to $\pi$-flux trapping, as seen in the center column of Fig.~\ref{SMfig:phase_diagram}.
In contrast, in the extreme limit that $|D_0| \gg |J_0|$, $50\%$ of the phase space traps $\pi$-flux, as shown in Fig.~\ref{SMfig:large_D0}(b).

 \section{Minimal energy configuration in three-grain network}
\label{appendix:minimal_energy_config}

In this section, we analyze the minimal energy configuration of the three-grain Josephson network described by Eq.~\eqref{free_energy.three_grain.real_D0} in the main text.

\subsection{Periodicity of Josephson free energy}

Firstly, we note the periodicity of the free energy.
By the loop constraint in Eq.~\eqref{loop_constraint.three_grain}, two pairs of relative angles $\beta_{nm}$ can change by $\pi$ and still satisfy $\beta_{12} + \beta_{23} + \beta_{31} = 0 \mod 2\pi$.
For example, under $\beta_{12} \rightarrow \beta_{12}+\pi$ and $\beta_{23} \rightarrow \beta_{23}$, it follows that $\beta_{31} \rightarrow \beta_{31}+\pi$.
Under $\beta_{nm} \rightarrow \beta_{nm} \pm \pi$, it follows that the effective Josephson coupling $\mathcal{J}_{nm}^\mathrm{eff}$ in Eq.~\eqref{effective_josephson_coupling} changes sign.
Hence, under $\beta_{12} \rightarrow \beta_{12}+\pi$ and $\beta_{31} \rightarrow \beta_{31}+\pi$, we have introduced an additional two $\pi$-junctions, which does not change the overall sign of the $\pi$-flux condition in Eq.~\eqref{pi-flux_condition}.
The $\pi$-periodicity in $d$ vector configurations can be seen in the phase diagrams in Fig.~\ref{SMfig:phase_diagram}.

Secondly, the system is insensitive to the sign of $d$ vectors.
For example, under $\beta_1 \rightarrow \beta_1 + \pi$, corresponding to $\hat{\mathbf{d}}_1 \rightarrow - \hat{\mathbf{d}}_1$, both $\beta_{12} \rightarrow \beta_{12} + \pi$ and $\beta_{31} \rightarrow \beta_{31}+\pi$.
This similarly introduces an even number of $\pi$-junctions, and for the same reason, the $\pi$-flux trapping condition is unchanged.
In other words, the free energy is agnostic to the sign of $d$ vectors, unlike the three-site frustration problem in classical magnetism.

\subsection{\texorpdfstring{$120^\circ$}{120}-ordered \texorpdfstring{$d$}{d} vector configuration}

Next, we determine the relative alignment of $d$ vectors in the three-grain junction geometry that minimizes the free energy in Eq.~\eqref{free_energy.three_grain.real_D0}.
The free energy is given by
\begin{equation}
    F_J = \sum_{\langle n, m \rangle} K \cos (\beta_{nm} + \delta)
    \cos \Delta \varphi_{nm},
\end{equation}
in which $K = \sqrt{J_0^2 + D_0^2}$.
To minimize each Josephson junction in the network, it follows that
\begin{equation}
    \cos \Delta \varphi_{nm} = - \mathrm{sgn} \cos (\beta_{nm} + \delta).
\end{equation}
Hence, to minimize, $\Delta \varphi_{nm} = \pi$ if the following condition is satisfied:
\begin{equation}
    -\frac{\pi}{2} \leq \beta_{nm} + \delta \leq \frac{\pi}{2}.
\end{equation}

In an ideal scenario, $\beta_{nm} = \pi - \delta$ for each bond, but the loop constraint in Eq.~\eqref{loop_constraint.three_grain} prevents the optimal configuration from being achieved.
Firstly, we assume that the overall $\mathrm{U}(1)$ phases will reorient to optimize the free energy, such that
\begin{equation}
    F_J = -K \sum_{\langle n, m \rangle}
    |\cos (\beta_{nm} + \delta)|.
\end{equation}
Let us define
\begin{math}
    x_{i} = \beta_{nm} + \delta,
\end{math}
which satisfies
\begin{math}
    x_1  +x_2 + x_3 = 3\delta,
\end{math}
from the loop constraint.
Here, $i$ refers to the bonds in the three-grain ring.
We find the extrema of
\begin{equation}
    f(x_1, x_2, x_3) = \sum_{i} |\cos x_i|,
\end{equation}
with the above constraint.
The problem can be solved using Lagrange multipliers,
\begin{equation}
    \mathcal{L}(x_i)
    =
    \sum_i |\cos x_i| + \lambda(x_1 + x_2 + x_3 - 3\delta).
\end{equation}
From the Euler-Lagrange equation, it follows that
\begin{equation}
    \lambda 
    = g(x_i)
    \equiv
    \sin (x_i) \mathrm{sgn} ( \cos x_i).
\end{equation}
The function $g(x_i)$ is monotonically increasing on intervals where $\mathrm{sgn}( \cos x_i)$ is constant.
The extrema occur for $g(x_1) = g(x_2) = g(x_3) = \lambda,$ corresponding to $\beta_{12} = \beta_{23} = \beta_{31} = \beta_0$, in which
\begin{equation}
    \beta_0
    =
    \frac{2\pi m}{3},
\end{equation}
for $m = 0, 1, 2$, following from the loop constraint.